\documentclass[twocolumn,showpacs,prb,amsmath,amssymb,floatfix]{revtex4}
\usepackage{amsmath,amssymb}
\usepackage{bm}
\usepackage{epsfig,psfrag}

\newcommand{\bd}{\bm}
\newcommand{\opvec}[1]{{\bd {#1}}}
\newcommand{\uvec}[1]{{\bd {#1}}}

\begin{document}
\title{Persistent spin currents in mesoscopic Haldane-gap spin rings}
  
\author{Florian Sch\"{u}tz, Marcus Kollar, and Peter Kopietz}
  
\affiliation{Institut f\"{u}r Theoretische Physik, Universit\"{a}t
  Frankfurt, Robert-Mayer-Strasse 8, 60054 Frankfurt, Germany}

\date{August 12, 2003}

  \begin{abstract}
    Using a modified spin-wave approach, we show that in the presence
    of an inhomogeneous magnetic field or an in-plane inhomogeneous
    electric field a mesoscopic antiferromagnetic Heisenberg ring with
    integer spin (i.e., a Haldane-gap system) exhibits a persistent
    circulating spin current.  Due to quantum fluctuations the current
    has a finite limit on the order of $(-g\mu_{\text{B}}) c /L$ at
    zero temperature, provided the staggered correlation length $\xi$
    exceeds the circumference $L$ of the ring, in close analogy to
    ballistic charge currents in mesoscopic normal-metal rings. Here
    $c$ is the spin-wave velocity, $g$ is the gyromagnetic ratio, and
    $\mu_{\text{B}}$ is the Bohr magneton.  For $\xi \ll L$ the current
    is exponentially suppressed.
  \end{abstract}
  
  
  \maketitle

  \section{Introduction}

  In a recent paper,\cite{psc} henceforth referred to as I, we have
  shown that the mesoscopic persistent current in a normal-metal ring
  threaded by an Aharonov-Bohm flux has an analog in spin rings:
  specifically, we have calculated the persistent spin current $I_s$
  in thermal equilibrium that circulates a ferromagnetic Heisenberg
  ring subject to a crown-shaped magnetic field with magnitude $|
  {\bd{B}}| $.  Within linear spin-wave theory we have shown that the
  associated magnetization current $I_m = ( g \mu_{\text{B}} / \hbar ) I_s$
  (where $g $ is the gyromagnetic ratio and $\mu_{\text{B}}$ is the Bohr
  magneton) is of the form\cite{footnote_convention}
  \begin{equation}
    I_{m} =  -\frac{g \mu_{\text{B}}}{ L} \sum_{k}
    \frac{v_k}{e^{ ( \epsilon_k + | {\bd{h}} |  )/ T } -1 }
    \, ,
    \label{eq:Issw}
  \end{equation} 
  where $L$ is the circumference of the ring, $\epsilon_k$ is the
  energy dispersion, and $v_k = \hbar^{-1} \partial \epsilon_k /
  \partial k$ is the velocity of the magnons, $ | {\bd{h}} | = g \mu_B
  | {\bd{B}} |$, and $T$ is the temperature.  In the presence of a
  crown-shaped inhomogeneous magnetic field the direction of the
  magnetization in the classical ground state covers a finite solid
  angle $\Omega_m$ on the unit sphere as we move once around the ring.
  Then the Bloch wave vectors are quantized as $k_n = \frac{2 \pi}{L} (
  n + \frac{\Omega_m}{2 \pi} )$, where in the continuum approximation
  the allowed values of $n$ are $n=0, \pm 1, \pm 2 , \ldots$.
  
\begin{figure}[tb]
    \centering \epsfig{file=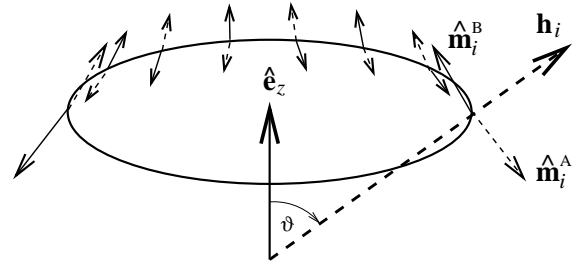,width=75mm}
    \caption{Classical ground state of an antiferromagnetic Heisenberg ring 
      in a crown-shaped magnetic field; $\uvec{m}_i^A$ and
      $\uvec{m}^B_i$ are the directions of the spins on sublattices
      $A$ and $B$.}
    \label{fig:crown_shape}
  \end{figure}

  The magnetization current (\ref{eq:Issw}) vanishes for $T
  \rightarrow 0$.  Physically, this is due to the fact that no quantum
  fluctuations are present in a ferromagnet, so that at $T=0$ there
  are no magnons to carry the spin current.  It is thus tempting to
  speculate that the analogous current for an antiferromagnetic
  Heisenberg ring will be finite even at $T=0$ due to quantum
  fluctuations.  In this work we shall show that this is indeed the
  case and present a quantitative calculation of the current in
  Haldane-gap antiferromagnets (i.e.,  antiferromagnetic Heisenberg
  rings with integer spin $S$) using a modified spin-wave theory.
  \cite{Takahashi87,mod_spin_wave,Kollar03} Our main result is that
  the ground state of a Haldane-gap spin ring subject to an
  inhomogeneous magnetic field supports a finite magnetization
  current, which in the limit where the staggered correlation length
  $\xi$ is large compared with $L$ has a sawtooth shape as a function
  of the geometric flux $\Omega$,
  \begin{equation}
    I_m = (-g \mu_{\text{B}}) \frac{c}{L} \left( 1 - \frac{\Omega}{\pi}
    \right)\,, 
    \qquad 
    0<\frac{\Omega}{2 \pi}< 1\,,
    \label{eq:sawtoothaf}
  \end{equation}
  where $c$ is the spin-wave velocity.  Here $\Omega$ is the solid
  angle on the unit sphere traced out be the local N\'eel vector as
  one moves once around the ring. Such a state can be produced by an
  inhomogeneous magnetic field, as depicted in
  Fig.~\ref{fig:crown_shape}.  Similar to the case of a ferromagnet
  discussed in I, the magnetization current is carried by magnons
  which are subject to mesoscopic interference due to the geometric
  phase associated with the inhomogeneous nature of the classical
  ground state. 
  Due to quantum fluctuations, the ring is endowed
  with an electric dipole moment even in the ground state.
  The absence of true long-range order in one-dimensional Heisenberg
  antiferromagnets leads for integer spin $S$ to a finite spin-correlation 
  length $\xi$ and a Haldane gap of the order of $\hbar c
  / \xi$ between the ground-state energy and the lowest triplet
  excitation. These features are correctly captured by a modified
  spin-wave theory\cite{Takahashi87,Kollar03} which we use here to
  calculate the spin current in Haldane-gap systems. Note that this
  approach is only appropriate for integer $S$, where the low-energy
  excitations can be viewed as renormalized spin waves. In contrast,
  for half integer $S$ the spectrum is gapless and spin correlations
  decay algebraically.  \cite{Auerbach94} The elementary excitations
  are then spinons, so that the modified spin-wave theory does not
  correctly reproduce the low-energy physics. In this case the
  effective low-energy theory is a Tomonaga-Luttinger model.  Recently
  Meier and Loss \cite{Meier03} used such a model to discuss the spin
  current in $S=1/2$ antiferromagnetic spin chains for a two-terminal
  geometry.
  
  While Eq. (\ref{eq:Issw}) is the bosonic analog of the persistent
  charge current \cite{Buettiker83,Imry97} in a normal one-dimensional
  metal ring threaded by an Aharonov-Bohm flux $\phi$, Eq.
  (\ref{eq:sawtoothaf}) is formally identical with the persistent
  charge current in a ballistic metal ring at zero temperature.
  Recall that for spinless fermions at constant chemical potential
  $\mu$ the charge current in the ballistic regime can be written as
  \cite{Cheung88}
  \begin{equation}
    I_c = \frac{-e}{L}\sum_k \frac{v_k}{ e^{ ( \epsilon_k - \mu )/T } + 1 } \,
   .
    \label{eq:I_c}
  \end{equation}
  Here $\epsilon_k$ is the energy dispersion and $v_k$ is the
  corresponding velocity of an electron in the state with wave number
  $k$.  Due to the Aharonov-Bohm flux the wave vectors are quantized as
  $k_n = \frac{2 \pi}{L} ( n - \phi /\phi_0 ) $, where $\phi_0 = h c /
  e $ is the flux quantum.  For free fermions at zero temperature Eq.
  (\ref{eq:I_c}) reduces to a sawtooth function.  For an even number
  of electrons at $T=0$ one obtains
  \begin{equation}
    I_c = (-e) \frac{ v_F}{L}\,\left(1 - \frac{2\phi}{\phi_0}\right)\,,
    \qquad 
    0<\frac{\phi}{\phi_0}< 1\,,
    \label{eq:sawtooth}
  \end{equation}
  where $v_F$ is the Fermi velocity.  With the replacements $e
  \rightarrow g \mu_{\text{B}}$, $v_F \rightarrow c$, and $\phi / \phi_0
  \rightarrow \Omega / 2 \pi$ the zero-temperature charge current
  (\ref{eq:sawtooth}) is formally identical with the zero-temperature
  magnetization current (\ref{eq:sawtoothaf}) in a Haldane-gap spin ring.
  Finite temperature, disorder, and phase-breaking scattering all have
  a similar effect on the persistent charge current in one dimension.
  \cite{Cheung88} For a weak perturbation, they smooth the
  discontinuity around $\phi=0$ and with increasing strength higher
  harmonics are exponentially suppressed such that a sinusoidal shape
  is approached.  In the limit of a very strong perturbation the
  current is exponentially suppressed with the relevant length or
  energy scale, i.e., the current becomes proportional to $\exp(- T /
  T^*)$, $\exp(- L/L_{\xi})$, or $\exp(-L/L_{\phi})$ under the
  influence of a nonzero temperature, strong disorder or strong
  inelastic scattering, respectively. Here $T^*$ is the temperature
  scale associated with the discrete level spacing, $L_{\xi}$ is the
  localization length, and $L_{\phi}$ is the phase-coherence length.
  Below we show that in the case of spin currents in Haldane gap
  systems the correlation length $\xi$ plays a similar role: for $\xi
  \ll L$ the magnetization current is exponentially suppressed and
  becomes sinusoidal.
  
  The rest of the paper is organized as follows. In
  Sec.~\ref{sec:spin_wave} we apply linear spin-wave theory to an
  antiferromagnet subject to an inhomogeneous magnetic field. After
  calculating the spectrum of magnon excitations, we obtain the spin
  current as a gauge-invariant derivative of the flux-dependent part
  of the free energy and discuss the result for $T=0$ in na\"{\i}ve
  spin-wave theory. To justify and generalize the spin-wave approach,
  we then apply a modified spin-wave theory, where the absence of 
  long-range order in a one-dimensional antiferromagnet is taken into
  account. In Sec.~\ref{sec:el_field} an antiferromagnet in an
  inhomogeneous {\it electric} field is considered.  In Appendix A,
  the classical ground state is obtained explicitly for the simple
  geometry of a crown-shaped field, and Appendix B contains some
  mathematical details of the calculation of the magnetization
  current.

  \section{Spin-wave theory}
  \label{sec:spin_wave}

  \subsection{Spectrum}
  \label{sec:spectrum}
  
  We start with a nearest-neighbor antiferromagnetic ($J > 0$)
  Heisenberg ring in an inhomogeneous magnetic field
  \begin{equation}
    H = \sum_{i=1}^N\left[ J{\bd S}_i\cdot{\bd S}_{i+1} -  
      {\bd h}_i\cdot{\bd S}_i\right]\,,
    \label{eq:H}
  \end{equation}
  where ${\bd S}_i$ are localized spin operators at sites ${ l}_i$ of
  a one-dimensional lattice with lattice spacing $a=L/N$, and ${\bd
    S}_i^2=S(S+1)$.  Periodic boundary conditions ${\bd S}_{N+1}={\bd
    S}_1$ are used, ${\bd h}_i=g\mu_{\text{B}}{\bd B}_i$, and $N$ is
  assumed to be even.  For a spin-wave expansion the classical ground
  state has to be determined by minimizing the energy in
  Eq.~(\ref{eq:H}) with ${\bd S}_i$ replaced by $S\hat{\bd m}_i$,
  where $\hat{\bd m}_i$ is a unit vector.  To consider fluctuations,
  the spin operators are then represented by bosonic creation and
  annihilation operators in the standard way,
  \begin{equation}
    \opvec{S}_i\cdot\uvec{m}_i = S-b_i^{\dag}b_i\,,\quad
    \opvec{S}_i\cdot\uvec{e}_i^+ = \sqrt{2S}b_i [1+\mathcal{O}(S^{-1})]
    \,,
  \end{equation}
  where $\uvec{e}_i^+=\uvec{e}^1_i+i\uvec{e}_i^2$, and $\uvec{e}_i^1$
  and $\uvec{e}_i^2$ are two unit vectors in the plane perpendicular
  to $\uvec{m}_i$, such that
  $\{\uvec{e}_i^1,\uvec{e}_i^2,\uvec{m}_i\}$ is a right-handed local
  basis in spin space. There is a local gauge freedom in the choice of
  the transverse basis vectors $\uvec{e}_i^1$ and $\uvec{e}_i^2$ which
  can be arbitrarily rotated around $\uvec{m}_i$. With the notation of
  I the spin-wave Hamiltonian then reads
  \begin{multline}
    H_{\text{sw}}= \frac{JS}2\sum_i\Big\{\big[
    (1+\uvec{m}_i\cdot\uvec{m}_{i+1}) e^{i(\omega_{i\rightarrow i+1} -
      \omega_{i+1\rightarrow i})}
    b_i^{\dag}b_{i+1}\\
    -(1-\uvec{m}_i\cdot\uvec{m}_{i+1}) e^{i(\omega_{i\rightarrow i+1}
      + \omega_{i+1\rightarrow i})} b_i^{\dag}b_{i+1}^{\dag}
    +\text{h.c.}\big]\\[2mm]
    -2\uvec{m}_i\cdot\uvec{m}_{i+1}(b_i^{\dag}b_i+b_{i+1}^{\dag}b_{i+1})
    +{\bd h}_i\cdot\uvec{m}_ib_i^{\dag}b_i\Big\}\,,
      \label{eq:H_sw}
  \end{multline}
  where $\omega_{i\rightarrow j}$ is the angle of rotation around
  $\uvec{m}_i$ that takes $\uvec{m}_i\times\uvec{m}_j$ into
  $\uvec{e}_i^1$. The transformation to another local right-handed
  triad, associated with site $i$ and the bond $(ij)$ according to
  $\{\tilde{\uvec{e}}_i^1,\tilde{\uvec{e}}_i^2=\uvec{m}_i\times\uvec{m}_j,\uvec{m}_i\}$,
  then reads
  \begin{equation}
    \uvec{e}_i^{\pm}=e^{\pm i\omega_{i\rightarrow j}}\tilde{\uvec{e}}_i^{\pm}\,.
  \end{equation}
  Equation~(\ref{eq:H_sw}) is manifestly invariant under the $\text{U}(1)$ gauge
  transformation $\omega_{i\rightarrow j}\rightarrow
  \omega_{i\rightarrow j} + \alpha_i$, $b_i\rightarrow
  e^{i\alpha_i}b_i$.  We will now assume a sufficiently large ring so
  that the direction of the magnetic field varies only slightly on the
  scale of the lattice spacing $a$. The classical ground state
  then locally resembles a N\'eel state and the local N\'eel vector
  $\uvec{n}_i=(-1)^{i+1}\uvec{m}_i$ varies smoothly as a function
  of position on the lattice and is oriented almost orthogonal to
  the local direction of the magnetic field. We thus have
  $\uvec{m}_i\cdot\uvec{m}_{i+1}=-1+O(1/N)$ and the terms
  involving the combination $ b_i^{\dag}b_{i+1}$ in
  Eq.~(\ref{eq:H_sw}) can be neglected to leading order in $1/N$.  In
  this approximation a local spin deviation follows the direction of
  the classical ground state as it moves around the ring. It picks up
  a geometrical phase leading to interference of its wave function in
  close analogy to Aharonov-Bohm interference in charge transport. The
  effect of the inhomogeneous field can be incorporated via the gauge
  transformation
  \begin{equation}
    \alpha_j=\sum_{i=1}^{j-1}(-1)^{j+i}
    (\omega_{i\rightarrow i+1} + \omega_{i+1\rightarrow i}).
  \end{equation}
  The new boundary conditions then become in the bosonic language
  \begin{equation}
    b_{i+N}=e^{\pm i\Omega}b_i\,,\quad
    \Omega = \alpha_{N+1}\,,
    \label{eq:boundary}
  \end{equation}
  where the upper/lower sign is valid for sublattice $A$/$B$ (odd/even
  $i$).  The resulting quadratic bosonic Hamiltonian is standard for
  an antiferromagnetic ring with nearest-neighbor interactions
  \begin{equation}
    H_{\text{AFM}}=\sum_i\big[-JS(b_i b_{i+1}+b_i^{\dag}b_{i+1}^{\dag})
    + (2JS+h_s)b_i^{\dag}b_i\big]\,,
    \label{eq:H_AFM}
  \end{equation}
  except for the twisted boundary condition (\ref{eq:boundary}).  An
  additional staggered field $h_s$ in the direction of the classical
  ground-state vectors has been introduced as a technical tool for the
  discussion in Sec.~\ref{sec:mod_spin_wave}.  $H_{\text{AFM}}$ is
  diagonalized as usual by first performing Fourier transformations with
  different signs on the two sublattices:
  \begin{equation}
    a_k=\sqrt{\frac2N}\sum_{i\in A} e^{-ikl_i}b_i\,,\quad
    b_k=\sqrt{\frac2N}\sum_{i\in B} e^{+ikl_i}b_i\,,
  \end{equation}
  where the allowed wave vectors are given by
  \begin{equation}
    k_n = \frac{2\pi}{L}\left(n+\frac{\Omega}{2\pi}\right)\,,
    \quad
    n=0,\dots,\frac{N}2-1\,.
  \end{equation}
  The diagonal form of $H_{\text{AFM}}$ is then achieved by the
  Bogoliubov transformation
  \begin{equation}
    \left(\begin{array}{c}a_k\\b_k^{\dag}\end{array}\right)
    =
    \left(\begin{array}{cc}\cosh\theta_k&\sinh\theta_k\\
      \sinh\theta_k&\cosh\theta_k\end{array}\right)
    \left(\begin{array}{c}\alpha_k\\\beta_k^{\dag}\end{array}\right)\,,
  \end{equation}
  with
  \begin{equation}
    \tanh(2\theta_k)=\frac{\cos(ka)}{1+\tilde{h}_s}\,,
    \quad
    \tilde{h}_s=h_s/2JS\,.
  \end{equation}
  The diagonal Hamiltonian contains constant terms due to quantum
  fluctuations,
  \begin{equation}
    H_{\text{AFM}}=\sum_k \epsilon_k (\alpha_k^{\dag}\alpha_k + \beta_k^{\dag}\beta_k +1) 
    -NJS(1+\tilde{h}_s)\,,
  \end{equation}
  where the quasiparticle energies are given by
  \begin{equation}
    \epsilon_k=2JS\sqrt{\Delta^2+\sin^2(ka)}\,,\quad
    \Delta^2=\tilde{h}_s(\tilde{h}_s+2)\,,
  \end{equation} 
  and the free energy is obtained from
  \begin{equation}
    F_{\text{sw}}(\Omega)=2T\sum_k\ln\left[2\sinh \frac{\epsilon_k}{2T}\right]-NJS(1+\tilde{h}_s)\,.
  \end{equation}
  Thus we have shown that to leading order in $1/N$ thermodynamic
  quantities depend on the inhomogeneity of the field only via the
  single phase $\Omega$. Geometrically $\Omega$ is the anholonomy
  associated with the parallel transport of a vector orthogonal to the
  local N\'eel vector,\cite{Shapere89} in analogy to the
  ferromagnetic case.  To see this more clearly,  we consider for each bond $(ij)$
  two additional sets of local right-handed triads containing
  the N\'eel vector $\uvec{n}_i$ instead of $\uvec{m}_i$. These triads
  are given by $\{\bar{\bd e}_i^1={\bd e}_i^1,\bar{\bd
    e}_i^2=(-1)^{i+1}{\bd e}_i^2,\uvec{n}_i\}$ and $\{\tilde{\bar{\bd
      e}}_i^1,\tilde{\bar{\bd
      e}}_i^2=\uvec{n}_i\times\uvec{n}_j,\uvec{n}_i\}$, and are related
  by a rotation around $\uvec{n}_i$. For the associated spherical
  vectors this reads
  \begin{equation}
    \bar{\bd e}_i^{\pm}=e^{\pm i\bar{\omega}_{i\rightarrow j}} \tilde{\bar{\bd e}}_i^{\pm},
  \end{equation}
  where the rotation angles $\bar{\omega}_{i\rightarrow j}$ are given
  by
  \begin{equation}
    \bar{\omega}_{i\rightarrow j}= i\pi + (-1)^{i+1}\omega_{i\rightarrow j}
    \qquad\text{for}\qquad j=i\pm1.
  \end{equation}
  We can now express $\Omega$ as
  \begin{equation}
    \Omega=\sum_{i=1}^N 
    (\bar{\omega}_{i\rightarrow i+1}-\bar{\omega}_{i+1\rightarrow i})\mod 2\pi\,,
  \end{equation}
  which is of the form obtained in I for the ferromagnet. $\Omega$ is
  thus the anholonomy of a vector orthogonal to the local N\'eel
  vector that is transported around the ring by discrete rotations
  around $\uvec{n}_i\times\uvec{n}_{i+1}$. Alternatively, a continuous
  parallel transport can be used around a path of geodesics connecting
  the unit vectors $\uvec{n}_i$ on the unit sphere. $\Omega$ is
  therefore equal to the solid angle subtended by this closed path of
  geodesics.

  \subsection{Magnetization current}
  \label{sec:magn_curr}
  
  In I we have shown that the $\text{U}(1)$ gauge symmetry associated with
  the choice of the local transverse basis is connected with a
  conserved current that was identified with the longitudinal
  component of the spin current. For an antiferromagnet the
  longitudinal spin current is conveniently defined in the direction
  of the local N\'eel vector and can be written as a gauge-invariant
  derivative of the free energy
  \begin{equation}
    I_s = \langle \uvec{n}_i\cdot {\bd I}_{i\rightarrow i+1}\rangle 
    = -(-1)^{i+1}\left\langle \frac{\partial H_{\text{sw}}}{\partial\omega_{i\rightarrow i+1}}\right\rangle
    = - \frac{\partial F_{\text{sw}}}{\partial\Omega}\,,\label{eq:I_s}
  \end{equation}
  where ${\bd I}_{i\rightarrow i+1}=J\opvec{S}_i\times\opvec{S}_{i+1}$
  is the spin current from site $i$ to $i+1$.  Similar to the
  ferromagnetic case, the presence of a longitudinal spin current can
  be understood semiclassically as follows. The local spin deviation
  from the classical ground state is essentially perpendicular to
  $\uvec{n}_i$ and varies slightly from $i$ to $i+1$. The spin
  deviations on neighboring sites are therefore in a plane that does
  not contain $\uvec{n}_i$, so that their cross product appearing in
  ${\bd I}_{i\rightarrow i+1}$ has a nonvanishing component $I_s$ in
  the direction of $\uvec{n}_i$.  This spin current $I_s$ corresponds
  to a current of magnetic dipoles that are locally oriented in the
  direction of the N\'eel vector $\uvec{n}_i$, which varies smoothly
  as we move along the ring. The spin current generates an electric
  dipole field which has the same form as discussed in I.  For the
  magnetization current we obtain
  \begin{equation}
    I_m = \frac{g\mu_{\text{B}}}{\hbar}I_s=-\frac{2g\mu_{\text{B}}}{L}\sum_k c_k
    \Big[n_k+\frac12\Big]\,,
    \label{eq:I_m}
  \end{equation}
  where $c_k= \hbar^{-1} \partial\epsilon_k/\partial k$ is the
  velocity of a magnon with wave vector $k$ and
  $n_k=1/[\exp(\epsilon_k/T)-1]$ is the Bose occupation factor. The
  extra factor of $2$ is due to the two degenerate spin-wave modes.
  Equation~(\ref{eq:I_m}) is the antiferromagnetic spin analog of
  Eq.~(\ref{eq:I_c}). It clearly shows that the magnetization current
  is carried by antiferromagnetic magnons, which at this level of
  approximation are the only quasiparticles available for transport.
  The current has a finite limit, even for vanishing Bose occupation
  factors, due to quantum fluctuations.  From Eq.~(\ref{eq:I_m}) the
  current is clearly seen to be a periodic function of $\Omega$, so
  that the finite momentum sum can be further evaluated using the
  Poisson summation formula.\cite{Cheung88} Some details of the
  calculation are given in Appendix~\ref{sec:mom_sums}.  For a
  vanishing staggered field in the zero-temperature limit we obtain
  the simple result announced in the Introduction,
  \begin{equation}
    I_m =
    I_m^0\left(1-\frac{\Omega}{\pi}\right)\,,
    \quad
    0<\frac{\Omega}{2\pi}<1\,.
    \label{eq:Im_hs0}
  \end{equation}
  Here $I_m^0=-g\mu_{\text{B}}c/L$ is the magnetization current
  carried by a single magnon with the spin-wave velocity $c
  =c_{k\rightarrow 0^+}$ at the center of the Brillouin zone.  The
  sawtooth shape (see the solid line in Fig.~\ref{fig:I_plot}) of the
  current in Eq.~(\ref{eq:Im_hs0}) is reminiscent of
  Eq.~(\ref{eq:sawtooth}) for charge transport.  Indeed, for $T=0$
  Eq.~(\ref{eq:I_m}) is formally equivalent to Eq.~(\ref{eq:I_c}) for
  charge transport when the Fermi edge is replaced by the lower edge
  of the magnon band.

  \subsection{Modified spin-wave theory}
  \label{sec:mod_spin_wave}
  \begin{figure}[tb]
    \centering
    \epsfig{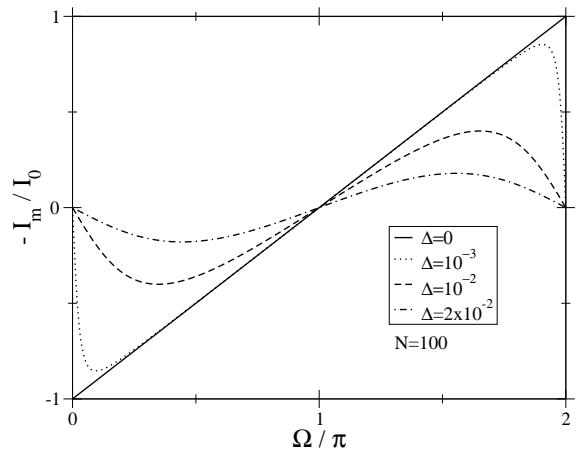}
    \caption{Magnetization current in a ring with 100 spins for 
      different values of the energy-gap parameter $\Delta$. The plots
      are produced by numerically evaluating Eq.~(\ref{eq:I_m}). For
      $\Delta=10^{-3}$ the curve is indistinguishable from the
      approximate expression in Eq.~(\ref{eq:I_small}), and
      Eq.~(\ref{eq:I_watson}) provides a good approximation for
      $\Delta=2\times10^{-2}$.}
    \label{fig:I_plot}
  \end{figure}
  The usual spin-wave theory employed so far is inconsistent when
  zero modes appear. Although the spin current remains finite, the
  sublattice magnetization diverges in the limit $\Omega\to0$.  This
  failure is related to the absence of long-range order in
  one-dimensional systems.  It can be resolved by a modified spin-wave
  theory which was first used by Takahashi for a one-dimensional
  ferromagnet \cite{Takahashi87} and then extended to various
  spin systems without long-range order, including antiferromagnets.
  \cite{mod_spin_wave} The constraint that is introduced in these
  theories was recently shown to follow naturally from a calculation
  at constant order parameter.\cite{Kollar03} In the present
  context, we introduce the additional constraint
  \begin{equation}
    \sum_i\langle\opvec{S}_i\cdot\uvec{m}_i\rangle = 0\,,
    \label{eq:constraint}
  \end{equation}
  which suppresses N\'eel order on average. This constraint is enforced
  via the staggered field $h_s$ in Eq.  (\ref{eq:H_AFM}) which acts as
  a Lagrange multiplier. The expectation value in Eq.
  (\ref{eq:constraint}) can be evaluated from $\partial
  F_{\text{sw}}/\partial h_s$, yielding the self-consistency condition
  \begin{equation}
    \frac2{N}\sum_k\frac{\partial\epsilon_k}{\partial h_s}\left[n_k+\frac12\right]=S+\frac12\,.
    \label{eq:h_s_self_con}
  \end{equation}
  Although the self-consistently determined $h_s$ is itself a periodic
  function of the geometric flux $\Omega$, the leading order for large
  $N$ is a constant and can be determined by replacing the sum in
  Eq.~(\ref{eq:h_s_self_con}) by an integral.  For $T=0$ the solution of
  Eq.~(\ref{eq:h_s_self_con}) yields the Haldane gap $2JS\Delta$,
  which is inversely proportional to the staggered correlation length
  $\xi$,\cite{Auerbach94}
  \begin{equation}
    \Delta=4\,e^{-\pi(S+1/2)}=\frac{a}{\sqrt{2}\xi}\,.
  \end{equation}
  The functional form of the magnetization current shows a crossover
  between the two qualitatively different regimes $\xi\gg L$ and
  $\xi\ll L$ (see Fig.~\ref{fig:I_plot}). In the former case
  $\Delta\ll 2\pi/N$ and at most one wave vector can be in the region
  $-\Delta<k<\Delta$ where the dispersion relation deviates strongly
  from the dispersion in the limit $\Delta=0$. When the contribution
  from this single wave vector is taken into account separately, we
  obtain
  \begin{equation}
    \frac{I_m}{I_m^0} = 
      \frac{\sin(2\Omega/N)}{2\sqrt{\Delta^2+\sin^2(\Omega/N)}}
      -\frac{\Omega}{\pi}\,,
      \quad
      -\pi<\Omega<\pi\,.
      \label{eq:I_small}
  \end{equation}
  In the case $\Delta=0$ this reduces to Eq.~(\ref{eq:Im_hs0}), provided
  $N\gg1$.  Thus, the effect of a finite $\Delta$ is to remove the
  discontinuity at $\Omega=0,2\pi$. On the other hand, in the limit
  $\xi\ll L$ many $k$ values are affected by the energy gap.  An
  analytic result for the spin current can nevertheless be derived as
  described in Appendix~\ref{sec:mom_sums}. We obtain the scaling form
  \begin{equation}
    \frac{I_m}{I_m^0} = \sqrt{\frac2{\pi}}
    \left(\frac{L}{\sqrt2\xi}\right)^{1/2}
    \exp\left(-\frac{L}{\sqrt2\xi}\right)\sin(\Omega)
    \,,
    \label{eq:I_watson}
  \end{equation}
  implying that the sinusoidal magnetization current is exponentially
  suppressed in the bulk limit, $L\gg\xi$.

  \section{Electric field}
  \label{sec:el_field}
  Moving magnetic dipoles represent an electric dipole moment
  \cite{Hirsch99} and are therefore affected by electric fields.  Due
  to this relativistic effect, which is essentially equivalent to
  spin-orbit coupling, the magnetic moments acquire an Aharonov-Casher
  phase.\cite{Aharonov84} For localized spin systems described by a
  Heisenberg Hamiltonian, the electric field can be taken into account
  phenomenologically by a substitution in the interaction term,
  \begin{equation}
    \opvec{S}_i\cdot\opvec{S}_j \longrightarrow
    \opvec{S}_i\cdot e^{{\bd \theta}_{ij}\times}\opvec{S}_j\,,
  \end{equation}
  as long as the electric field varies only weakly on the scale of the
  lattice spacing. Here,
  \begin{equation}
    {\bd\theta}_{ij} = \frac{g\mu_{\text{B}}}{\hbar c^2}
    \int_{l_i}^{l_j}d{\bd l}\times{\bd E}({\bd l})\,,
  \end{equation}
  and $e^{{\bd \theta}\times}$ denotes the SO(3) rotation matrix
  acting on a vector ${\bd m}$ according to\cite{Chandra90}
  \begin{equation}
    e^{{\bd \theta}\times}{\bd m}=\hat{\uvec{\theta}}(\hat{{\bd\theta}}\cdot{\bd m})+
    (\hat{\uvec{\theta}}\times{\bd m})\sin\theta-
    \hat{\uvec{\theta}}\times(\hat{\uvec{\theta}}\times{\bd m})\cos\theta\,,
  \end{equation}
  with ${\bd \theta}=\theta\hat{\bd \theta}$.
  For ferromagnetic coupling, inhomogeneous electric fields can lead
  to persistent magnetization currents \cite{Cao97} and a spin
  analog of the Hall effect was also shown to exist in electric
  fields.\cite{Meier03}
  \begin{figure}[tb]
    \centering
    \epsfig{file=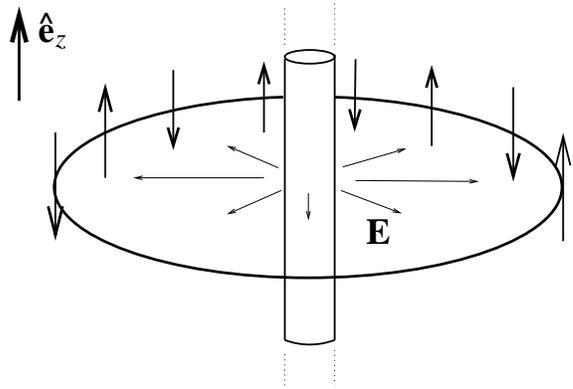,width=75mm}
    \caption{Antiferromagnetic Heisenberg ring in the electric field 
      produced by a line charge.}
    \label{fig:E_field}
  \end{figure}

  We now consider the antiferromagnetic ring in an electric field in
  the $x$-$y$ plane, e.g. produced by a charged line in the $z$ direction
  (see Fig.~\ref{fig:E_field}).  The rotation vectors
  ${\bd\theta}_{i,i+1}=\theta_{i,i+1}\uvec{e}^z$ are then all parallel
  to the $z$ axis and the Hamiltonian for vanishing magnetic field
  becomes
  \begin{equation}
    H = J\sum_i\left[\frac12(e^{i\theta_{i,i+1}}S_i^+S_{i+1}^- 
        + \text{h.c.}) + S_i^zS_{i+1}^z\right]\,.
  \end{equation}
  The classical ground state is easily shown to be a doubly degenerate
  N\'eel state with $\uvec{m}_i=\pm\uvec{e}^z$. The spin-wave
  expansion is thus straightforward. If a gauge transformation is used
  to eliminate the phase factors, we again obtain the standard bosonic
  Hamiltonian $H_{\text{AFM}}$ of Eq.~(\ref{eq:H_AFM}) with the
  boundary condition (\ref{eq:boundary}), where $\Omega$ is replaced
  by the total Aharonov-Casher phase
  \begin{equation}
    \Omega_{\text{AC}}=\sum_i \theta_{i,i+1} = 
    \frac{g\mu_{\text{B}}}{\hbar c^2}\oint d{\bd l}\cdot[\uvec{e}^z\times{\bd E}({\bd l})]\,.
  \end{equation}
  The spin current then only has a $z$ component which can be written as
  a gauge-invariant derivative of the free energy,
  \begin{equation}
    I_m = -\frac{\partial F_{\text{AFM}}}{\partial\Omega_{\text{AC}}}\,.
  \end{equation}
  This leads again to Eq.~(\ref{eq:I_m}) with $\Omega$ replaced by
  $\Omega_{\text{AC}}$ and all the results derived in the previous
  sections are also applicable in this context.
  
  It is also interesting to note that the situation of a radial
  electric field with $\theta_{i,i+1}=\frac{2\pi}N$ and an additional
  homogeneous magnetic field tilted with respect to the $z$ axis can be
  formally mapped onto a crown-shaped magnetic field alone via the
  transformation
  \begin{equation}
    \opvec{S}_i^{'}=e^{\frac{2\pi}N\uvec{e}^z\times}\opvec{S}_i\,.
  \end{equation}
  It would therefore be interesting to further investigate the
  combined effect of arbitrary inhomogeneous magnetic and electric
  fields on the produced spin currents to find situations that could
  be realized more easily in the laboratory for a possible
  experimental detection of the effect.

  \section{Summary and Outlook}
 
  In the last two decades, a lot of theoretical work focused on
  persistent equilibrium currents in mesoscopic normal-metal
  rings.\cite{Buettiker83,Imry97,Eckern02} In the 1990s, the
  experimental difficulties for the detection of the currents were
  overcome and an oscillating magnetization as a function of the
  magnetic flux was observed under various
  conditions.\cite{Levy90,Chandrasekhar91,Mailly93,Reulet95,Rabaud01,Jariwala01}
  In the ballistic regime, the main features of the experiment
  \cite{Mailly93} can be understood within a simple model of free
  fermions,\cite{Cheung88} but in the diffusive regime a generally
  accepted explanation of the experiments
  \cite{Levy90,Chandrasekhar91} is still lacking.
  
  In this work, we have shown that persistent magnetization currents
  are present in antiferromagnetic Heisenberg rings in inhomogeneous
  magnetic fields as well as in a radial electric field. Quantum
  fluctuations lead to ground-state currents, but fluctuations in low
  dimensions also produce exponential damping when the circumference
  $L$ of the ring becomes larger than the staggered correlation
  length. We have obtained explicit expressions for the current at
  $T=0$ in the two limits $L\ll\xi$ and $L\gg\xi$ within a modified
  spin-wave approach valid for integer spins, i.e., for Haldane-gap
  systems. The determination of the current for half-integer spin
  rings remains an interesting open problem.  Also, we have only
  considered clean systems in this work, i.e., we have focused on the
  ballistic regime.  Since for persistent charge current disorder is
  known to be important, it would also be very interesting to consider
  persistent magnetization currents in the diffusive regime of
  disordered magnets.
  
  In the past few years a new field of research (``spintronics'') has
  emerged, where the spin degree of freedom is used as a medium to
  transport information.\cite{Awschalom02}  For technical
  applications (such as quantum computation) it is important to control
  the quantum coherence in mesoscopic spin systems.  The persistent
  spin current discussed in this work can be viewed as a quantitative
  measure for the degree of quantum coherence in the ground state of
  the system.  Similar to the ferromagnetic case discussed in I, the
  magnetization current circulating a Heisenberg antiferromagnet
  generates an electric dipole field.  Due to screening effects, the
  corresponding voltage drop may be rather difficult to detect
  experimentally. However, in view of the rapid development of the
  field of spintronics, it does not seem unreasonable to expect that
  before the end of this decade new experimental techniques will be
  available to detect persistent spin currents in Heisenberg rings.

  \vspace{7mm}
 
  We thank Florian Meier and Bernd Wolf for discussions.  This work
  was supported by the DFG via Forschergruppe FOR 412, Project No. KO
  1442/5-1.

  \appendix

  \section{Crown-shaped magnetic field}
  \label{sec:crown_shape}
  \begin{figure}[t]
    \centering \epsfig{file=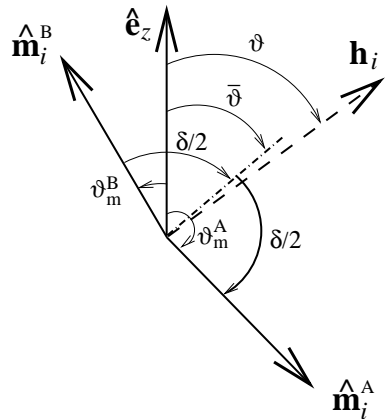,width=50mm}
    \caption{Definition of angles for the description of the classical
      ground state of an antiferromagnetic Heisenberg ring in a
      crown-shaped magnetic field.}
    \label{fig:def_angle}
  \end{figure}
  Here we determine the classical ground state of an antiferromagnetic
  ring in the crown-shaped magnetic field depicted in
  Fig.~\ref{fig:crown_shape}, i.e., a field with
  \begin{equation}
    {\bd h}_i= h [ \sin(\vartheta)\cos(2\pi l_i/L),
    \sin(\vartheta)\sin(2\pi l_i/L),\cos(\vartheta) ]
  \end{equation}
  For very strong magnetic fields the classical unit vectors
  $\uvec{m}_i$ will be aligned parallel to the field and the ground
  state will have the full rotational symmetry of the applied field.
  Below a critical spin-flip field $h_c(\vartheta)$ it will be
  energetically favorable to form two sublattices with different
  angles $\vartheta_m^{A/B}$ to the $z$-axis (see Fig.
  \ref{fig:def_angle}). Introducing the relative and average angles
  \begin{equation}
    \delta=\frac12(\vartheta_m^A-\vartheta_m^B)\qquad  
    \bar{\vartheta}=\frac12(\vartheta_m^A+\vartheta_m^B)\,,
  \end{equation}
  a minimum of the classical energy is reached for
  \begin{eqnarray}
    \sin(\vartheta-\bar{\vartheta})\cos(\delta)
    &=&-\frac{J S}{h}\,\epsilon_-\sin(2\bar{\vartheta})
    \label{eq:angle}
    \\
    \cos(\vartheta-\bar{\vartheta})\sin(\delta)
    &=&+\frac{J S}{h}\,\epsilon_+\sin(2\delta)\,,
    \label{eq:delta}
  \end{eqnarray}
  where we have defined $\epsilon_{\pm}=1\pm\cos(2\pi/N)$.
  For very strong magnetic fields $h>h_c$, we have $\delta=0$ and
  Eq.~(\ref{eq:angle}) reduces to its ferromagnetic analogue (see Eq.
  (14) in I).  For $\delta\neq0$ the two equations can be combined to
  give
  \begin{equation}
    \sin(2(\vartheta-\bar{\vartheta}))=
    -\left(\frac{2JS}h\right)^2\sin^2(2\pi/N)\sin(2\bar{\vartheta})\,.
    \label{eq:theta_bar}
  \end{equation}
  Thus for large rings the magnetic field $h\sim JS/N$ necessary to
  produce an inhomogeneous classical ground state is well below the
  spin-flip field $h_c\sim JS$. For $h\sim JS/N \ll JS$ we have
  $\delta\sim\pi/2$ and the classical ground state locally resembles a
  N\'eel state as assumed in Sec.~\ref{sec:spin_wave}.

  \section{Calculation of spin current}
  \label{sec:mom_sums}  
  In this appendix we present some details of the calculation of the
  functional form of the spin current at zero temperature. Eq.
  (\ref{eq:I_m}) can be written in the form
  \begin{equation}
    \frac{I_m}{I_m^0}=\sum_{n=0}^{N/2-1} 
    f\!\left(\frac{2\pi}{N}\left(n+\frac{\Omega}{2\pi}\right)\right)\,,
  \end{equation}
  where $f(k)=d/d(ka)(\epsilon_k/2JS)$ is a periodic function of its
  argument with $f(x+\pi)=f(x)=-f(-x)$. Consequently $I_m$ is a
  periodic function of $\Omega$ and we can proceed by calculating its
  Fourier coefficients $C_{\nu}$ in
  \begin{equation}
    \frac{I_m}{I_m^0}=\sum_{\nu=-\infty}^{\infty}C_{\nu}e^{i\nu\Omega}\,.
  \end{equation}
  By appropriate substitution in the corresponding integral one
  eliminates the finite momentum sum and obtains
  \begin{equation}
    C_{\nu} = -\frac{\nu N^2\Delta}{2i}\,g_{\nu N}\!\left(-\Delta^{-2}\right)\,,
  \end{equation}
  where $g_{2l+1}$ is zero and $g_{2l}$ is given by
  \begin{eqnarray}
    g_{2l}(z)&=&\int_0^{2\pi}\frac{d\omega}{2\pi}
    \sqrt{1-z\sin^2\omega}\,\,e^{-i2l\omega}\nonumber\\
    &=&\binom{\frac12}{l}\left(\frac{z}4\right)^l
    \ _2F_1\!\left(l-\frac12,l+\frac12;2l+1;z\right)\nonumber\\
    &=& (-1)^l\sqrt{1-z}\,g_{2l}\!\left(\frac{z}{z-1}\right)\,.
    \label{eq:hyper}
  \end{eqnarray}
  Using an asymptotic expansion of the hypergeometric function
  \cite{Watson18} for large $l$, we derive an expression for the
  spin current for large system sizes. In particular, the condition
  that the leading term in the expansion be sufficient can be shown to
  be equivalent to $L/\xi\gg\mathcal{O}(1)$.  After some algebra we
  obtain Eq.~(\ref{eq:I_watson}).
  


\begin{thebibliography}{99}
\bibitem{footnote_convention}%
   The extra minus sign in Eq.~(\ref{eq:Issw}) in comparison to
   Eq.~(15) of Ref.~\onlinecite{psc} is due to a different sign
   convention in the Fourier transformation used in the present work.
\end{thebibliography}
\end{document}